\documentclass[conference,10pt]{IEEEtran}
\usepackage{cite}
\usepackage{array}
\usepackage{amssymb,amsmath}		
\usepackage{graphics}		
\usepackage{longtable}          
\usepackage{subfigure}
\usepackage{float}
\usepackage{amssymb,amsmath}

\newcommand{\vectornorm}[1]{{\left|\left|#1\right|\right|}_{2}}

\begin{document}
\title{Bayesian Quantized Network Coding \\ via Belief Propagation}
\author{\IEEEauthorblockN{Mahdy Nabaee and Fabrice Labeau}
\IEEEauthorblockA{Electrical and Computer Engineering Department, McGill University, Montreal, QC}
}

\maketitle

\begin{abstract}
In this paper, we propose an alternative for routing based packet forwarding, which uses network coding to increase transmission efficiency, in terms of both compression and error resilience.
This non-adaptive encoding is called quantized network coding, which involves random linear mapping in the real field, followed by quantization to cope with the finite capacity of the links.
At the gateway node, which collects received quantized network coder packets, minimum mean squared error decoding is performed, by using belief propagation in the factor graph representation.
Our simulation results show a significant improvement, in terms of the number of required packets to recover the messages, which can be interpreted as an embedded distributed source coding for correlated messages.

\end{abstract}
\IEEEpeerreviewmaketitle

\begin{keywords}
Network coding, Bayesian compressed sensing, belief propagation, minimum mean squared error estimation.
\end{keywords}

\section{Introduction}
\label{sec:Intro}

Data gathering in sensor networks has drawn attention to network coding \cite{fragouli2009network} as an alternative for routing based packet forwarding \cite{al2004routing} because of its flexibility, and robustness to the network changes and link failures. 
In the case of correlated messages, performing network coding on top of distributed source coding \cite{xiong2004distributed} is shown to be optimal, in terms of the achievable information rates \cite{NCCorr_NIFwithCOrrSo}.
However, appropriate encoding rates have to be known at the encoders, which requires transmission of an overhead and affects  the flexibility and distributed nature of sensor networks.

Recently, the possibility of adopting \textit{non-adaptive} joint source network coding has been studied by using the concepts of compressed sensing \cite{CS} and sparse recovery \cite{rabbat,CdataGathering,feizi2011power,luo2011compressive,naba1,bassi2012compressive}.
As a first major work to formulate and investigate theoretical feasibility of compressed sensing based network coding, we proposed to use \textit{Quantized Network Coding} (QNC) \cite{naba1,naba2}, which involves random linear network coding in the real field and quantization. 
In \cite{naba1}, our decoding scheme was based on $\ell_1$-minimization (using linear programming \cite{1542412}), which is shown to be optimal for recovery of exactly sparse messages, from noiseless measurements.
In this paper, we study optimal Minimum Mean Squared Error (MMSE) decoding and feasibility of its implementation in practical cases. 
Specifically, we propose to use a near optimal MMSE decoding based on Belief Propagation (BP).

Quantized network coding using low density coefficients is described and formulated in section~\ref{sec:QNC}.
We discuss optimal MMSE decoding for our QNC scenario with known priori in section~\ref{sec:MMSEdec}.
In section~\ref{sec:BPdec}, we describe our BP based MMSE decoding, followed by our simulation results in section~\ref{sec:simRes}. Finally, our concluding remarks are presented in section~\ref{sec:Conclusions}.

\section{Quantized Network Coding}
\label{sec:QNC}

Consider a network (graph), $\mathcal{G}=(\mathcal{V},\mathcal{E})$, with the set of nodes $\mathcal{V}=\{1,\ldots,n\}$, and the set of directed edges (links) $\mathcal{E}=\{1,\ldots,|\mathcal{E}|\}$.
Each edge, $e$, can maintain a lossless transmission from $tail(e)$ to $head(e)$, at a maximum rate of $C_e$ bits per use.
The input content of edge $e$ (which is the same as its output content), at time index $t$, is represented by $y_e(t)$, and is from a finite alphabet of size $2^{L C_e}$, where $L$ is the block length, transmitted in each time slot, between $t-1$ and $t$.
For each node, $v$, we define sets of incoming edges, $\textit{In}(v)=\{e: head(e)=v, e \in \mathcal{E}\}$, and outgoing edges, $\textit{Out}(v)=\{e: tail(e)=v, e \in \mathcal{E}\}$.
Moreover, each node $v$ has a random information source, $X_v$, where there is a transform matrix, $\phi_{n \times n}$, such that $\underline{X}=[X_v:v \in \mathcal{V}]$ and $\underline{S}=\phi^T  \underline{X}$ is $k$-sparse.
As in the rest of this paper, where we represent the realizations of random variables with lower case letters, the outcome realization of $\underline{X}$ is represented by $\underline{x}$.
In this paper, we study the (single session) data gathering, where all the messages, $X_v$'s, are to be transmitted to a single node, called decoder (or gateway), $v_0 \in \mathcal{V}$.

We defined QNC at each node, $v \in \mathcal{V}$, as follows \cite{naba1}:
\begin{equation}\label{Eq:QNC1}
Y_e(t)= \textbf{Q}_e \Big [\sum_{e' \in \textit{In}(v)} \beta_{e,e'}(t)\cdot Y_e(t-1)+\alpha_{e,v}(t)\cdot X_v \Big ], 
\end{equation}
where $\textbf{Q}_e[ \centerdot]$ is the quantizer (designed based on the value of $C_e$ and $L$, and the distribution of incoming contents and messages), associated with the outgoing edge $e \in \textit{Out}(v)$. The corresponding network coding coefficients, $\beta_{e,e'}(t)$ and $\alpha_{e,v}(t)$ are selected from real numbers: $\beta_{e,e'}(t),\alpha_{e,v}(t) \in \mathbb{R}$, and satisfy the normalizing condition of (3) in \cite{naba1}.
Initial rest condition is also assumed to be satisfied in our QNC scenario: $Y_e(1)=0,~\forall~e \in \mathcal{E}.$

Denoting the quantization noise at edge $e$ by $N_e(t)$, we have:
\begin{equation}
Y_e(t)= \sum_{e' \in \textit{In}(v)} \beta_{e,e'}(t)\cdot Y_e(t-1)+\alpha_{e,v}(t)\cdot X_v + N_e(t).
\end{equation}
This is equivalent to:
\begin{equation}\label{eq:linear}
\underline{Y}(t)=F(t) \cdot \underline{Y}(t-1)+A(t) \cdot \underline{X}+\underline{N}(t),
\end{equation}
where $\underline{N}(t)=[N_e(t):e \in \mathcal{E}]$, and:
\begin{equation}
 F(t)_{|\mathcal{E}|\times |\mathcal{E}|}: \{F(t)\}_{e,e'}=\left\{
\begin{array}{l l}
  \beta_{e,e'}(t)  & ,~\scriptsize tail(e)=head(e') \\
  0  &  ,~\mbox{otherwise} \\ \end{array} \right. , \nonumber
\end{equation}
\begin{equation}\label{Eq:defineAt}
 A(t)_{|\mathcal{E}|\times |\mathcal{V}|}: \{A(t)\}_{e,v}=\left\{
\begin{array}{l l}
  \alpha_{e,v}(t)  & ,~tail(e)=v \\
  0  &  ,~\mbox{otherwise} \\ \end{array} \right. . \nonumber
\end{equation}

Representing the marginal measurements (received packets to the decoder) at time $t$, by $\underline{Z}(t)$, we have: 
\begin{equation}
\underline{Z}(t)=[Y_e(t):e \in \textit{In}(v_0)]=B \cdot \underline{Y}(t),
\end{equation}
where:
\begin{equation}
\{B\}_{i,e}=\left\{
\begin{array}{l l}
  1  & ,~i~\mbox{corresponds to}~e,~e \in \textit{In}(v_0) \\
  0  &  ,~\mbox{otherwise} \\ \end{array} \right. . \nonumber
\end{equation}
We store marginal measurements, over time, and build up a \textit{total measurements vector}, $\underline{Z}_{\rm{tot}}(t)$:
\begin{equation}\label{Eq:totMeasEq}
\underline{Z}_{\rm{tot}}(t)=\left[ {\begin{array}{*{20}c}
	\underline{Z}(2) \\	
   \vdots   \\
   \underline{Z}(t)   \\
 \end{array} } \right]_{m \times 1},~~~m=(t-1) |\textit{In}(v_0)|.
\end{equation}
As a result of linearity of QNC scenario (Eq.~\ref{eq:linear}), we have \cite{naba1}:
\begin{equation}\label{Eq:measEq}
\underline{Z}_{\rm{tot}}(t)= \Psi_{\rm{tot}}(t) \cdot \underline{X} + \underline{N}_{\rm{eff,tot}}(t),
\end{equation}
where $\Psi_{\rm{tot}}(t)$ and $\underline{N}_{\rm{eff,tot}}(t)$ are called \textit{total measurement matrix} and \textit{total effective noise vector}, respectively.

In \cite{naba1}, a compressed sensing (\textit{i.e.} $\ell_1$-min) decoding is used to reconstruct $\underline{X}$ from noisy under-determined measurements, $\{\underline{Z}_{\rm{tot}}(t)\}$'s. 
Being able to recover $n$ different values, $X_v$'s, from $m$ measurements, where $m$ is usually much less than $n$, can be interpreted as an embedded distributed compression of inter-node correlated $X_v$'s.
Although this is feasible with respect to some distortion, the proposed $\ell_1$-min decoding does not offer an optimal solution, especially when a prior on $\underline{X}$ is available and has more information than sparsity.
In this paper, we address the optimal MMSE decoding in a Bayesian QNC scenario by studying the computational complexity of implementing such decoder.
Motivated by the work in \cite{baron2010bayesian}, near optimal implementation of MMSE decoding based on belief propagation and the appropriate design of network coding coefficients are discussed.

\section{Minimum Mean Square Error Decoding}
\label{sec:MMSEdec}

The a priori model used to characterize the messages is a Gaussian mixture model.
Specifically, we consider states, $Q_v$'s ($S_v$ is the $v$'th element of vector $\underline{S}$), corresponding to $S_v$'s, which are independent binary random variables with:
\begin{equation}
\textbf{P}(Q_v=1)=\frac{k}{n},~\forall v \in \mathcal{V}.
\end{equation}
Each state, $Q_v$, determines if $S_v$ is zero or not:
\begin{eqnarray}
Q_v=1 & \rightarrow & S_v \sim \mathcal{N}(0,\sigma^2_s), \nonumber\\
Q_v=0 &\rightarrow & S_v=0.
\end{eqnarray}
Therefore, a priori of independently modelled ${S}_v$'s is as follows:
\begin{equation}\label{Eq:priori}
\textbf{p}_{S_v}(s_v)=\frac{k}{n}\frac{1}{\sqrt{2\pi \sigma^2_s}} e^{-\frac{1}{2\sigma^2_s}s^2_i}
+(1-\frac{k}{n})  \mathbf{\delta}(s_v),
\end{equation}
where $\mathbf{\delta}(\centerdot)$ is Dirac delta function, and it also implies: 
\begin{equation}
\textbf{E}[X^2_v]=\textbf{E}[S^2_v]=\frac{k}{n}\sigma^2_s,~\forall v.
\end{equation}

To facilitate the use of notations, we define $\Omega_{e,v}(t)$ to be the transform coefficient from $X_v$ to $Y_e(t)$, describing the transfer of information of messages through the network.
When the quantization noises, $N_e(t)$'s, have small variance compared to that of the signal, $\textbf{E}[\vectornorm{\underline{X}}^2]$, the variance of $Y_e(t)$'s can be approximated with the variance of noiseless propagated information, that is:
\begin{equation}
\textbf{E}[Y_e^2(t)] \simeq \frac{k}{n} \sigma^2_s \sum_{v=1}^{n} \Omega^2_{e,v}(t).
\end{equation}
Moreover, $N_e(t)$'s are approximately independent and their variance, $\textbf{E}[N^2_e(t)]$, is proportional with the variance of corresponding quantizer input, $\textbf{E}[Y_e^2(t)]$. Hence:
\begin{equation}
\sigma^2_{e}(t)=\textbf{E}[N^2_e(t)] \simeq \frac{k}{n} \gamma_e(t) \sigma^2_s \sum_{v=1}^{n} \Omega^2_{e,v}(t),
\end{equation}
where $\gamma_e(t)$ is a positive scalar, depending on the quantizer design.
Defining 
\begin{equation}\label{Eq:Ntot}
\underline{N}_{\rm{tot}}(t)=\left[ {\begin{array}{*{20}c}
	\underline{N}(2) \\	
   \vdots   \\
   \underline{N}(t)   \\
 \end{array} } \right]_{(t-1)|\mathcal{E}| \times 1},
\end{equation}
the effective total measurement noise, $\underline{N}_{\rm{eff,tot}}(t)$, can be formulated according to:
\begin{equation}
\underline{N}_{\rm{eff,tot}}(t)=\Psi_{\rm{N,tot}}(t) \cdot \underline{N}_{\rm{tot}}(t).
\end{equation}
This implies:
\begin{equation}\label{Eq:En1}
\textbf{E}[\underline{N}_{\rm{eff,tot}}(t)  \underline{N}^T_{\rm{eff,tot}}(t)] =
\Psi_{\rm{N,tot}}(t)  \Lambda_{Q}(t)  \Psi^T_{\rm{N,tot}}(t).
\end{equation}
where $\Lambda_{Q}(t)$ is the diagonal covariance matrix of quantization noises:
\begin{equation}
\Lambda_{Q}(t)=\textbf{E}[\underline{N}_{\rm{tot}}(t) \cdot \underline{N}_{\rm{tot}}(t)^T].
\end{equation}

The MMSE estimation of $\underline{X}$ is calculated according to:
\begin{eqnarray}
\underline{\hat{x}}_{\rm{opt}}(t)&=&\textbf{E}\Big[\underline{X}\Big|\underline{Z}_{\rm{tot}}(t)=\underline{z}_{\rm{tot}}(t)\Big] \label{Eq:MMSEest1}  \\
&=&\phi~  \textbf{E}\Big[\underline{S}\Big|\underline{Z}_{\rm{tot}}(t)=\underline{z}_{\rm{tot}}(t)\Big] \nonumber \\
&=& \phi  \int_{-\underline{\infty}}^{+\underline{\infty}} \underline{s}~\textbf{p}_{\underline{S}}\Big(\underline{s}\Big|\underline{Z}_{\rm{tot}}(t)=\underline{z}_{\rm{tot}}(t)\Big) \cdot d\underline{s}, \label{Eq:MMSEest2} \nonumber
\end{eqnarray}
where:
\begin{equation}
\textbf{p}_{\underline{S}}\Big(\underline{s}\Big|\underline{Z}_{\rm{tot}}(t)=\underline{z}_{\rm{tot}}(t)\Big)=
\frac{\textbf{p}_{\underline{S}}(\underline{s})~ \textbf{p}_{\underline{Z}_{\rm{tot}}(t)}\Big(\underline{z}_{\rm{tot}}(t)\Big|\underline{S}=\underline{s}\Big)}{\textbf{p}_{\underline{Z}_{\rm{tot}}(t)}\Big(\underline{z}_{\rm{tot}}(t)\Big)}, \nonumber
\end{equation}
and,
\begin{equation}
\textbf{p}_{\underline{Z}_{\rm{tot}}(t)}\Big(\underline{z}_{\rm{tot}}(t)\Big)=\int_{-\underline{\infty}}^{+\underline{\infty}}
\textbf{p}_{\underline{Z}_{\rm{tot}}(t)}\Big(\underline{z}_{\rm{tot}}(t)\Big|\underline{S}=\underline{s}\Big) \textbf{p}_{\underline{S}}(\underline{s}) \cdot d\underline{s}. \nonumber
\end{equation}
Now, having a prior of $\underline{S}$ (Eq.~\ref{Eq:priori}), the distribution of quantization noises, and the measurement equation of (\ref{Eq:measEq}), one could calculate the posterior probability of $\underline{X}$ and its MMSE estimation, $\underline{\hat{x}}_{\rm{opt}}(t)$.
However, this requires a high computational complexity for the decoder, which makes it practically infeasible.
To tackle this issue, near optimal MMSE decoding by using Belief Propagation (BP) is proposed \cite{baron2010bayesian}.
Such decoders are based on sum product algorithm \cite{kschischang2001factor}, which is widely used in the literature of low density parity check codes.
In section~\ref{sec:BPdec}, we describe the BP based near optimal MMSE decoder, used to recover messages in the considered Bayesian framework.

\section{MMSE Decoding via Belief Propagation}
\label{sec:BPdec}

Belief propagation\footnote{in some cases also referred as message passing} is used to calculate an approximate version of posterior probability, where a low density factor graph representation of random linear measurements is available \cite{baron2010bayesian}. In \cite{DBLP:journals/corr/abs-1001-2228}, BP decoding is extended to recover from random linear measurements, even when the graph representation is dense.

Consider the QNC measurement equation of (\ref{Eq:measEq}) where the elements of total effective noise, $\{\underline{N}_{\rm{eff,tot}}(t)\}_i$'s, are dependent. By eigen decomposition of their covariance matrix,
\begin{equation}
\textbf{E}[\underline{N}_{\rm{eff,tot}}(t)  \underline{N}^T_{\rm{eff,tot}}(t)] = U_N(t) \cdot \Lambda_N(t) \cdot U_N^T(t),
\end{equation}
we define:
\begin{equation}
\underline{Z}'_{\rm{tot}}(t)= \Lambda^{-\frac{1}{2}}_N(t) U_N^T(t) \cdot \underline{Z}_{\rm{tot}}(t),
\end{equation}
and
\begin{equation}
\underline{N}'_{\rm{eff,tot}}(t)=\Lambda^{-\frac{1}{2}}_N(t) U_N^T(t) \cdot \underline{N}_{\rm{eff,tot}}(t),
\end{equation}
for which we have:
\begin{equation}\label{Eq:linearMeas2}
\underline{Z}'_{\rm{tot}}(t)=\Theta'_{\rm{tot}}(t) \cdot \underline{S}+\underline{N}'_{\rm{eff,tot}}(t).
\end{equation}
In \ref{Eq:linearMeas2}, $\Theta'_{\rm{tot}}(t)$ is as follows:
\begin{equation}
\Theta'_{\rm{tot}}(t)=\Lambda^{-\frac{1}{2}}_N(t) U^T_N(t) \cdot \Psi_{\rm{tot}}(t) \cdot \phi,
\end{equation}
and $\{\underline{N}'_{\rm{eff,tot}}(t)\}_i$'s are uncorrelated with unit variance.
We also assume the marginal quantization noises, $N_e(t)$'s, can fit into a Gaussian distribution.
As a result of this assumption, $\{\underline{N}'_{\rm{eff,tot}}(t)\}_i$'s are independent zero mean Gaussian random variables with unit variance.

\begin{center}
\begin{figure}[t]
\centering
\resizebox{.35\textwidth}{!}{
\includegraphics{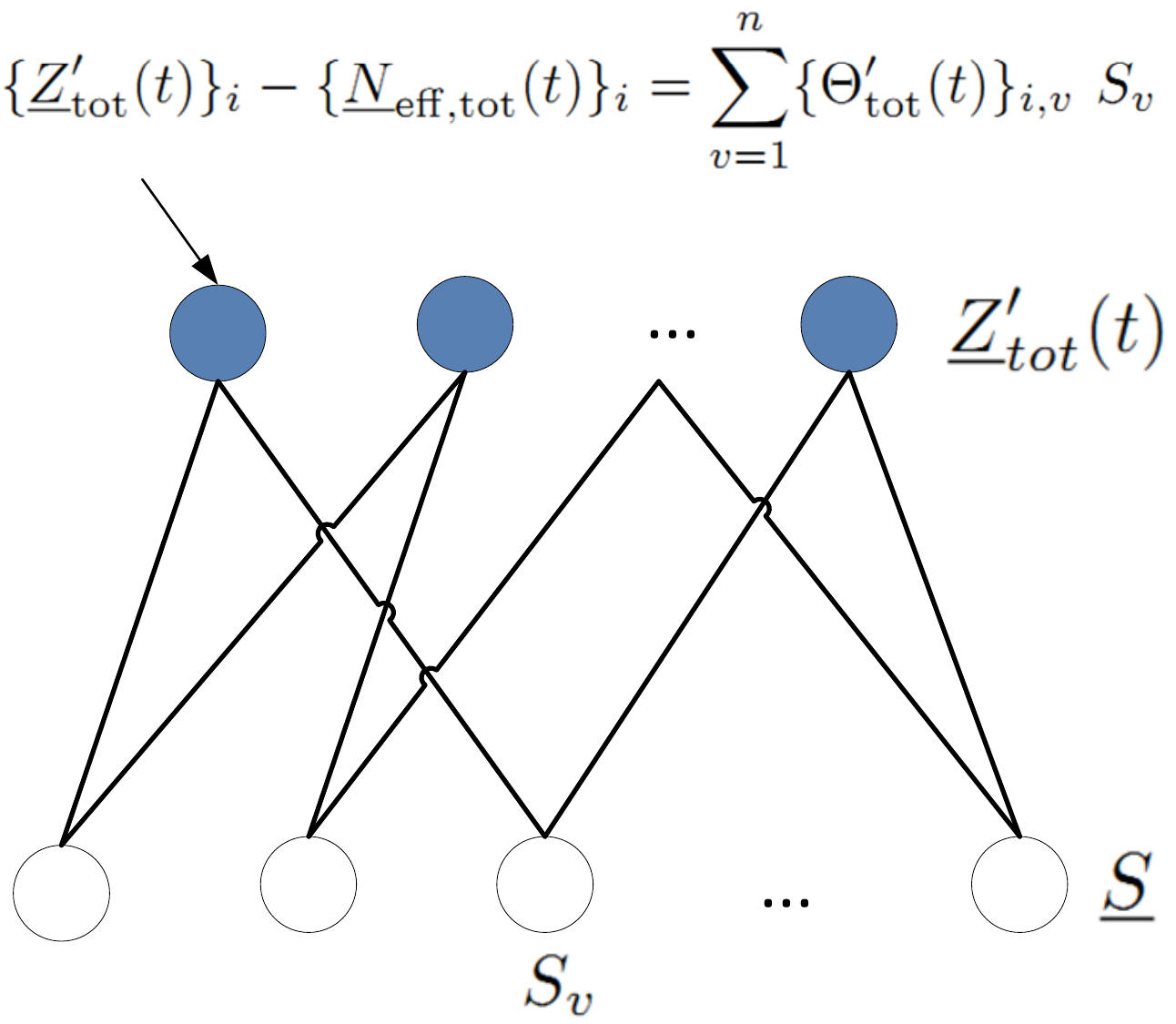}}
\caption{QNC can be represented by using a factor graph from the sparse messages to the noisy measurements.\label{fig:graphBP}
}
\end{figure}
\end{center}

The equivalent linear measurement equation of (\ref{Eq:linearMeas2}), which characterizes the QNC scenario, can be represented by a factor graph, as shown in Fig.~\ref{fig:graphBP}.
In this graph, each constraint node, $v$, $1 \leq v \leq n$, (gray node) is connected to a subset of variable nodes (white nodes), $i$, $1 \leq i \leq m$, for which $\{\Theta'_{\rm{tot}}(t)\}_{i,v} \neq 0$.
After enough passings of the beliefs, between the nodes of the factor graph, an approximate version of posterior probability of $S_v$'s may be obtained \cite{baron2010bayesian,DBLP:journals/corr/abs-1001-2228}.

In the following, we describe BP based decoding for our Bayesian QNC scenario:
\begin{enumerate}
\item The variable nodes have their prior information, \emph{i.e.} $\textbf{p}_{S_v}(\centerdot)$, as an initial belief to start with.
Explicitly, node $v$ sends this probability density function (PDF), $\textbf{p}_{v \rightarrow i}^{1}(\centerdot)=\textbf{p}_{S_v}(\centerdot)$, to its neighbour constraint nodes, $neib(v)=\{i':1 \leq i' \leq m, \{\Theta'_{\rm{tot}}(t)\}_{i',v} \neq 0\}$.

\item The received beliefs at the constraint node $i$, as well as the corresponding measurement, ${\underline{Z}_{\rm{tot}}(t)}_i={\underline{z}_{\rm{tot}}(t)}_i$, are used to calculate a backward belief. 
Specifically, for each $v$, where $v \in neib(i)=\{v':1 \leq v' \leq n, \{\Theta'_{\rm{tot}}(t)\}_{i,v'} \neq 0\}$, (\ref{Eq:BP1}) leads us to the update equation in (\ref{Eq:BP2}).\footnote{In BP update stage, the incoming beliefs (messages) to a node are assumed to be independent.}
In (\ref{Eq:BP2}), $\textbf{p}_{N'}(\centerdot)$ is the PDF of a zero mean Gaussian random variable with unit variance; $\star$ and $\tau$ represent the convolution operator, and the iteration index, respectively.

\begin{figure*}[t]
\begin{eqnarray}
\{\underline{z}'_{\rm{tot}}(t)\}_i & = & \{\underline{Z}'_{\rm{tot}}(t)\}_i \nonumber \\
&=& \sum_{v' \neq v,~ v' \in neib(i) } \{\Theta'_{\rm{tot}}(t)\}_{i,v'} S_{v'} + \{\Theta'_{\rm{tot}}(t)\}_{i,v} S_{v}+\{\underline{N}'_{\rm{eff,tot}}(t)\}_{i}
\label{Eq:BP1} \\
\textbf{p}_{i \rightarrow v}^{\tau}(\frac{\{\underline{z}'_{\rm{tot}}(t)\}_i}{\{\Theta'_{\rm{tot}}(t)\}_{i,v}}-s_v) &=&
\textbf{p}_{v'_{1} \rightarrow i}^{\tau}(\frac{\{\Theta'_{\rm{tot}}(t)\}_{i,v}}{\{\Theta'_{\rm{tot}}(t)\}_{i,v'_{1}}} s_v) ~\star ~ ... ~\star ~ \textbf{p}_{v'_{r} \rightarrow i}^{\tau}(\frac{\{\Theta'_{\rm{tot}}(t)\}_{i,v}}{\{\Theta'_{\rm{tot}}(t)\}_{i,v'_{r}}} s_v) ~\star ~  \textbf{p}_{N'}(\frac{s_v}{\{\Theta'_{\rm{tot}}(t)\}_{i,v}}), \nonumber \\
& & v'_1,\ldots v'_{r} \in neib(i) \setminus \{v\}
\label{Eq:BP2}
\end{eqnarray}
\end{figure*}

\item At the variable node $v$, given the received backward beliefs from the neighbour nodes, $\textbf{p}_{i \rightarrow v}^{\tau}(\centerdot)$, and a priori of $S_v$, the forward beliefs are updated according to:
\begin{equation}
\textbf{p}_{v \rightarrow i}^{\tau+1}(s_v)=c \cdot \textbf{p}_{S_i}(s_v) \cdot \prod_{v'} \textbf{p}_{i \rightarrow v'}^{\tau}(s_v),
\end{equation}
where $c$ is a constant, assuring the unit integral of $\textbf{p}_{v \rightarrow i}^{\tau+1}(\centerdot)$.
Given the posterior probabilities, one may calculate the BP based MMSE estimate of $S_v$'s:
\begin{eqnarray}
\hat{S}^\tau_v &= & \textbf{E}^{\tau}[S_v|\underline{Z}_{tot}(t)=\underline{z}_{tot}(t)] \nonumber\\
&=&\int_{-\infty}^{+\infty} s_v~ \textbf{p}_{v \rightarrow i}^{\tau}(s_v) \cdot d s_v,
\end{eqnarray}
and the corresponding $\underline{\hat{X}}^{\tau}=\phi \cdot \underline{\hat{S}}^{\tau}$, as an approximation for $\underline{\hat{X}}_{\rm{opt}}(t)$.
\item This procedure is repeated by going back to step 2 until some convergence criterion, such as:
\begin{equation}
\vectornorm{\hat{\underline{X}}^{\tau}-\hat{\underline{X}}^{\tau-1}} \leq \epsilon_{\rm{rec}},
\end{equation}
is met (where $\epsilon_{\rm{rec}}$ controls the precision of decoding).
\end{enumerate}

\section{Simulation Results}
\label{sec:simRes}

\begin{figure*}[t!]
\centering
\subfigure[$\frac{|\mathcal{E}|}{|\mathcal{V}|}=4$]{
\resizebox{.46\textwidth}{!}{
\includegraphics{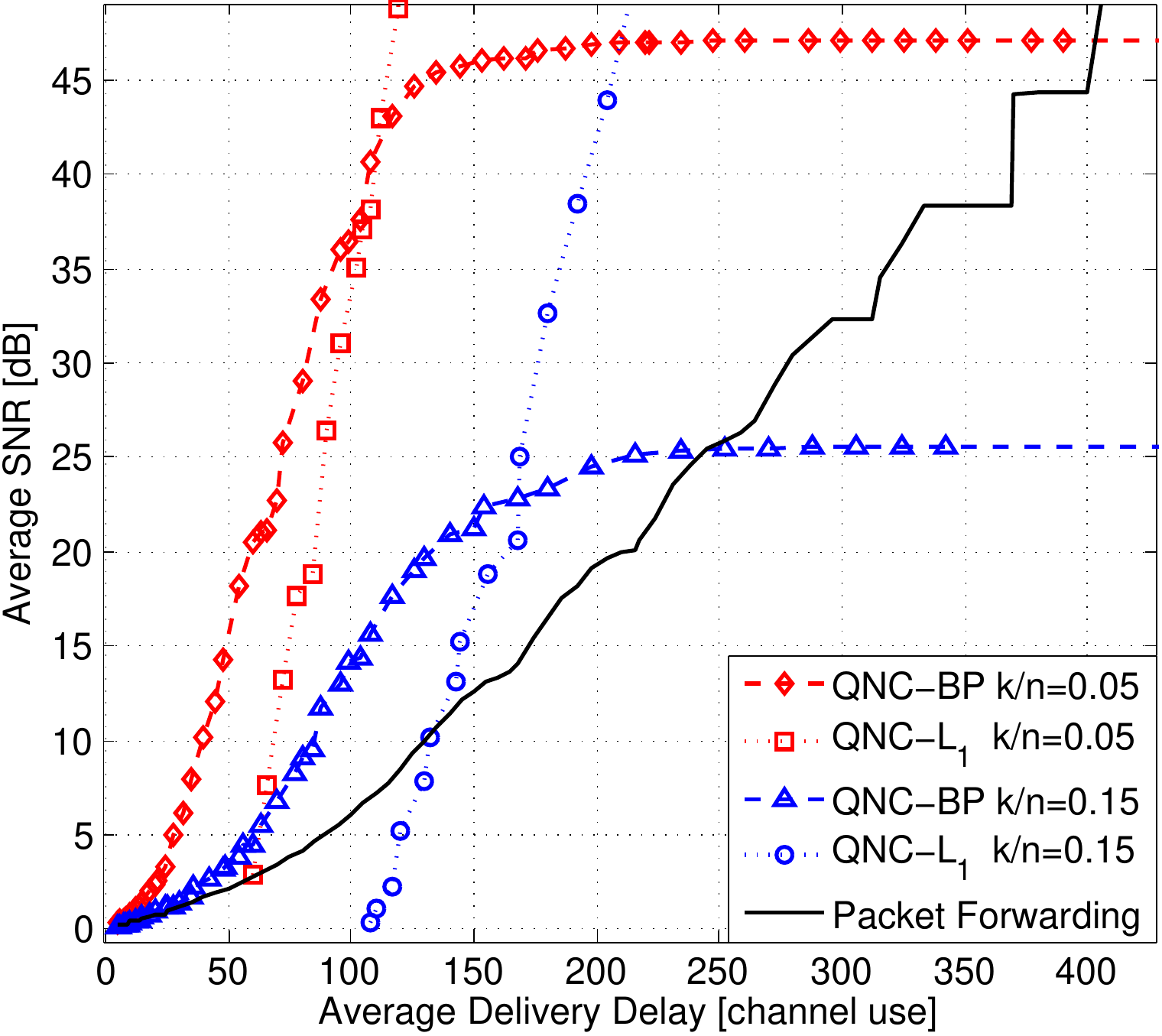}}
\label{fig:Dense4}
} \qquad
\subfigure[$\frac{|\mathcal{E}|}{|\mathcal{V}|}=8$]{
\resizebox{.46\textwidth}{!}{
\includegraphics{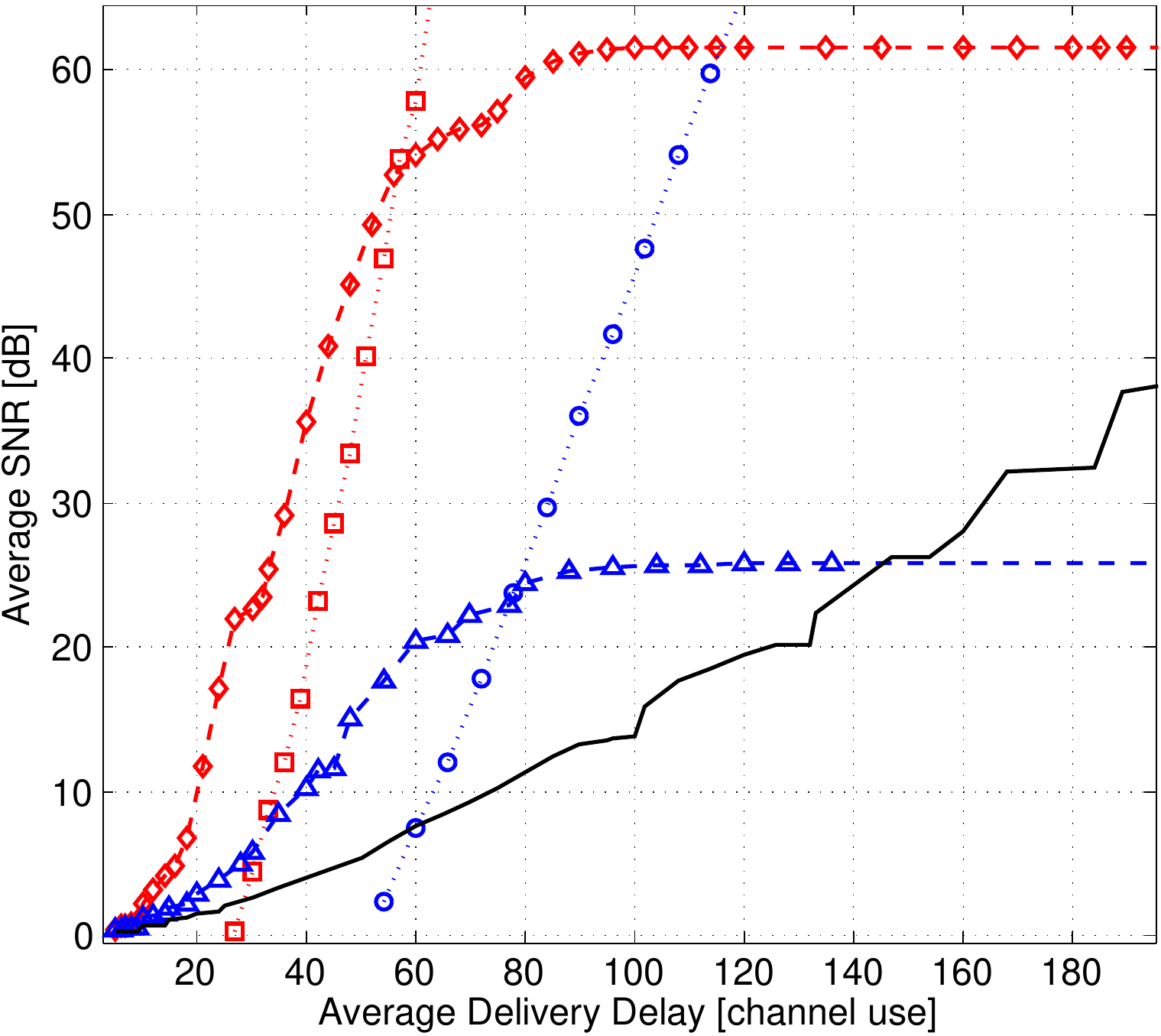}}
\label{fig:Dense8}
} \\
\subfigure[$\frac{|\mathcal{E}|}{|\mathcal{V}|}=12$]{
\resizebox{.46\textwidth}{!}{
\includegraphics{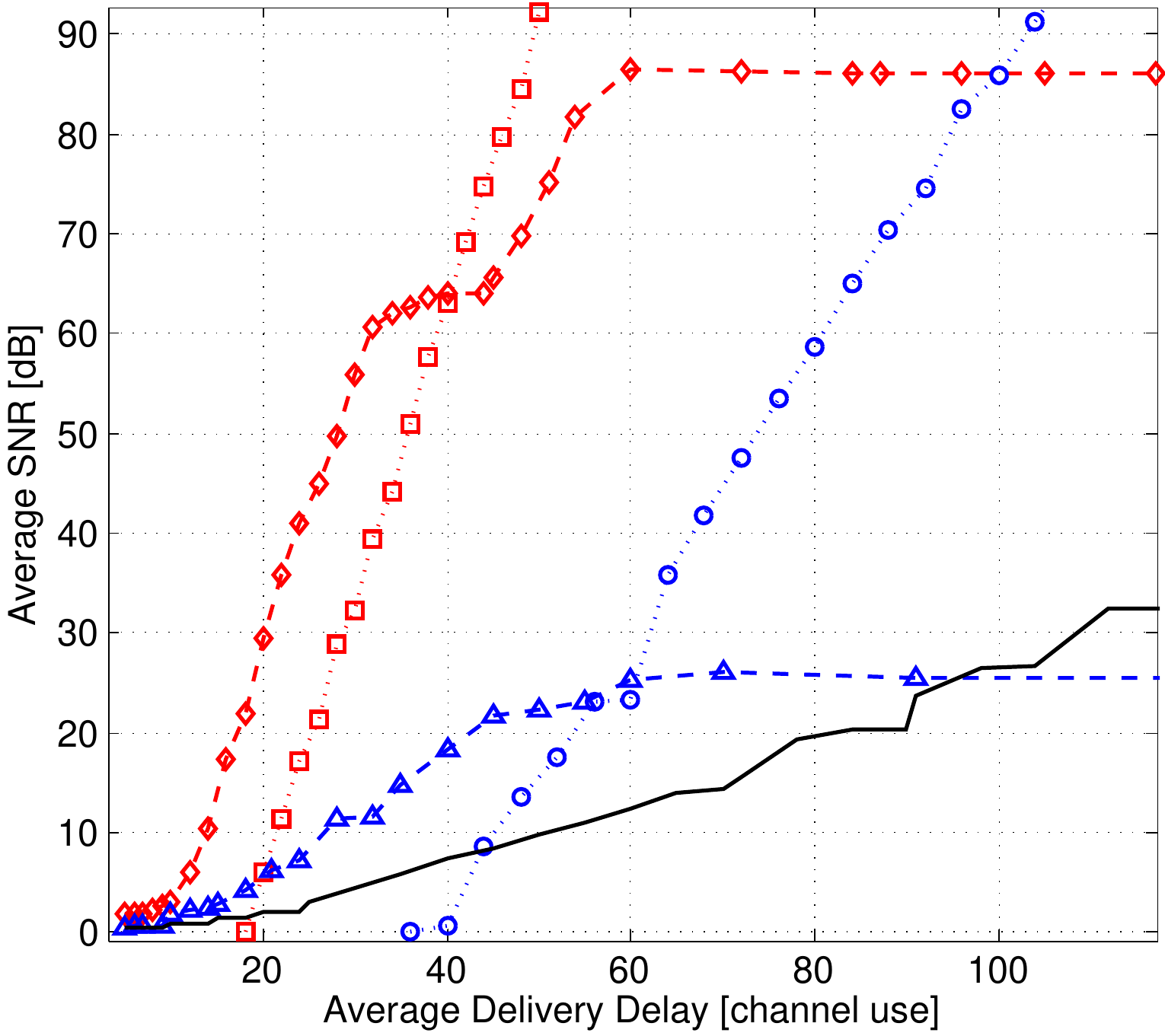}}
\label{fig:Dense12}
}
\caption{SNR versus delivery delay of QNC and Packet Forwarding for different sparsity factors and edge densities, using dense measurement matrices\label{fig:subfigureExampleDense}.
}
\end{figure*}

In this section, we evaluate the performance of the proposed QNC by comparing it with the conventional routing based packet forwarding.
Specifically, we generate random deployments of a network with $n=100$ nodes and $|\mathcal{E}|=400,800,1200$ uniformly distributed edges, where one of the nodes is randomly picked to be the gateway (decoder) node.
For each deployment, we also randomly generate $\underline{S}$'s, with a mixture Gaussian distribution, as described in (\ref{Eq:priori}). 
The messages are derived from different sparsity factors of $\frac{k}{n}=0.05,0.15$, and $\sigma^2_s=5$.\footnote{Note that the value of $\sigma^2_s$ does not affect the results, since it will scale the variance of quantization noises too.} Furthermore, the sparsifying matrix, $\phi$, is randomly generated and orthonormal.

For each deployment, we run QNC with different block lengths, $L$, and decoded the received packets at the decoder node to obtain $\underline{\hat{x}}(t)$.
The network coding coefficients, $\alpha_{e,v}(t)$'s and $\beta_{e,e'}(t)$'s, used in QNC scenario, are generated such that the resulting $\Psi_{tot}(t)$ is a dense Gaussian matrix.
Specifically, $\alpha_{e,v}(2)$'s are derived from independent zero mean Gaussian distributions, and the rest of $\alpha_{e,v}(t)$'s, $t>2$, are set to zero. Moreover, $\beta_{e,e'}(t)$'s are chosen to be locally orthonormal, as described explicitly in theorem~3.1 in \cite{naba1}.
BP based MMSE decoding as well as $\ell_1$-min decoding are used to reconstruct the messages.
The BP based decoder is as described in section~\ref{sec:BPdec}, and is implemented by using the implementation in \cite{DBLP:journals/corr/abs-1001-2228}.
$\ell_1$-min decoding is described in \cite{naba1}, theorem~4.1, and uses the open source optimization toolbox in \cite{cvx1}.

For each deployment, we also simulate packet forwarding to transmit messages to the decoder node. The route used to forward the packets is optimized (in terms of delivery delay) and calculated using the Dijkstra algorithm \cite{dijkstra1959note}. 
Continuous value messages are also quantized at the source nodes, by using a uniform quantizer with $2^{L C_e}$ levels.

For each SNR (quality), the best choice of block length, $L$, is found for both QNC and packet forwarding scenarios and used to minimize the corresponding delivery delay.
We present the results, by averaging them over different realizations of network deployments and messages.
In Fig.~\ref{fig:subfigureExampleDense}, the resulting average SNR is depicted versus the average delivery delay, obtained for different sparsity factors and density of edges, using dense measurement matrices.

As shown in Figs.~\ref{fig:Dense4},\ref{fig:Dense8},\ref{fig:Dense12}, the performance of using QNC is better than that of using routing based packet forwarding.
Specifically, the adopted $\ell_1$-min decoder, proposed in \cite{naba1}, already outperforms packet forwarding for all SNR values.
Using BP based MMSE decoding helps us improve the performance for some (especially low) SNR values, compared to $\ell_1$-min decoding.
Moreover, as it is expected, when the sparsity factor, $\frac{k}{n}$, of messages increases (meaning higher correlation between $X_v$'s), the gap between the QNC and packet forwarding curves increases. 
However, there is a drawback in using BP decoder for some SNR values, especially when the sparsity factor of messages, $\frac{k}{n}$, is high (\textit{i.e.} there is not a high correlation between $X_v$'s).
Such cases can be explained to be a result of propagation of quantization noises, in the network, which increases the noise power in the measurements.

\section{Conclusions}
\label{sec:Conclusions}
We have made improvements in the throughput of sensor networks, by introducing a network coding based approach for transmission of correlated sensed data to a gateway node.
Conventional linear network coding is joined with the concepts of Bayesian compressed sensing to efficiently embed distributed source coding in network coding.
On the other hand, belief propagation has helped us to discuss on near optimal decoding of quantized network coded messages, while computational resource constraints were intended to be met.
Our simulation results show significant savings for QNC in terms of delivery delay, when compared with conventional packet forwarding.
Moreover, using the proposed BP based MMSE decoder for QNC scenario helped us to require a smaller delivery delay, for (relatively) low SNR values.
As a lacking point in the studies of BP based decoding, we are still working to derive theoretical bounds on the performance of our decoder. This would give us a better understanding about the optimality of adopted decoder, with respect to the infinite block length information theoretic bounds.

\section*{Acknowledgment}
This work was supported by Hydro-Québec, the Natural Sciences and Engineering Research Council of Canada and McGill University in the framework of the NSERC/Hydro-Québec/McGill Industrial Research Chair in Interactive Information Infrastructure for the Power Grid.

\bibliographystyle{ieeetr}
\bibliography{RefarXiv}
\end{document}